\documentclass{ws-procs9x6}

\begin{document}

\title{From Parity Violation to Hadron Structure and more . . .}

\author{D.~H. Beck}

\address{Department of Physics, \\
University of Illinois at Urbana-Champaign, \\
1110 West Green Street, \\ 
Urbana, IL USA 61801-3080\\ 
E-mail: dhbeck@uiuc.edu}


\maketitle

\abstracts{
New developments in physics associated with parity-violating interactions are
discussed in this PAVI 2002 workshop summary.}

\section{Introduction}

The parity-violating property of the weak interaction allows us to observe its effects
in many situations where it would otherwise be practically impossible.  This property in turn
affords many opportunities to 
study the weak interaction itself, and, perhaps
surprisingly, to study hadronic structure as well.  For those of us in the field it
is easy to forget that significant advances are being made - the experiments and the
calculations look quite different than they did 5 - 10 years ago; the PAVI 2002 workshop 
provides a timely snapshot of our recent history and present activities.  This talk is
divided into three sections -- advances in technology, in understanding of the physics and the
open questions for the next meeting.

How do the measurements and calculations described below change the way we think about
physics -- the gold standard test for any endeavor in our field?  
Although the answer is perhaps more obvious for the investigations of the
weak interaction, in my view, the advances in hadronic structure are of equal importance.
There are lots of ways to express our ignorance of the manifestations of QCD in hadronic
structure; suffice to say that we have a pretty poor picture of how more than 99\% of 
visible mass in the universe is put together.

\section{Technology}

Unlike most electron scattering experiments, parity-violation measurements are very
tightly coupled to beam properties.  The quality of the polarized beam and measurements
of its properties are crucial.

There has been tremendous recent progress in the overall figure of merit for these 
experiments 
\begin{equation}
\frac{\Delta A}{A}= \frac{1}{f.o.m.} = \frac{1}{A_{phys}P_e N},
\end{equation}
where $N$ is the number of measured particles, 
because of the advent of high polarization ($P_e$) electron sources.  Whereas experiments 
were previously limited to polarizations of less than 50\% for bulk GaAs, new strained
crystals in use at Mainz,\cite{ref:Aulenbacher} SLAC\cite{ref:Humensky} and JLab\cite{ref:Baylac} have provided high currents with polarizations in excess of 80\%.  These crystal
materials continue to be improved, now involving multilayer structures,
 yielding larger and more uniform strains with quantum
efficiencies of order 1\%.  By activating only the central region of the crystal with
Cs, emission from the crystal edges (which can hit the vacuum chamber) is greatly reduced
and the crystal lifetime is consequently increased.\cite{ref:Farkhondeh}  Not surprisingly, the lasers being
used for these sources also continue to improve, both in terms of power and stability.

These strained crystals necessarily have an analyzing power for linearly polarized light
(caused by the strain) and because the circular polarization of the incident laser light
is not perfect, care must be taken to reduce the effects of the residual linearly
polarized component.  This is particularly difficult at SLAC where the laser beam diameter
is large (14 mm) and the associated large diameter Pockels cell generates only 99.2\%
circular polarization.  Nevertheless, the helicity-correlated position differences have
been reduced to the few tens of nm.  The intensity asymmetries are now quite routinely
corrected to the ppm level, most often using a scheme originated at Bates where a second
Pockels cell (the first being used to produce the circular polarization of the light)
is used as a variable shutter to adjust the laser intensity in a helicity-correlated 
manner.  More sophisticated analysis techniques are now being used to better understand
the transport of the various components of the polarized light in an effort to further
reduce these effects.

The uncertainty in beam polarization continues to be one of, or the largest, systematic uncertainty
in these experiments.  As the precision of the actual parity-violating asymmetry determinations
improves, beam polarimetry must keep up.  Combinations of conventional Mott and M\"oller 
measurements\cite{ref:Grames}, measurements using Compton backscattering\cite{ref:Burtin} and
developments of new analyzers\cite{ref:Collin} are supporting the new experiments.  My sense,
however, is that improvements in these measurements will have to be pursued even more aggressively
in the future.

With the success of early measurements of parity-violating electron scattering have come
more ambitious experiments with more sophisticated technology.  In order to measure\cite{ref:Pitt}
$Q_{weak}$ at JLab a roughly 2 kW liquid hydrogen target is required (with very stringent
limits on boiling).  In order to help determine accurately the neutron radius in lead,\cite{ref:Michaels} a
new JLab experiment will use a diamond backed lead target to deal with the heating from
intense beam required.  Detectors have also become more sophisticated.  The G0 experiment\cite{ref:Furget}
at JLab will, for the first time, utilize a ``mixed'' detector in a parity-violation 
measurement with the addition of Cerenkov detectors to supplement the scintillation
detectors in the backward angle phase for the purpose of pion rejection.  The detectors
for the E158 experiment\cite{ref:Souder} at SLAC have the unusual requirement that they must be radiation
hard because of the enormous M\"oller rate.  Both G0 and PVA4\cite{ref:Maas} at Mainz are, for the first
time in such experiments, counting individual particles and therefore 
using large numbers of detectors with the associated (standard for other experiments)
difficulties of maintenance of thresholds, gains, etc.  G0 also has sophisticated
custom electronics to produce histograms in hardware to allow total count rates of order
100 MHz.  PVA4 uses advanced trigger electronics to reject physics backgrounds (inelastic
electrons) at very high real rates ($\sim$ 100 kHz per channel, $\sim$ 100 MHz overall).

\section {Physics}

It is the parity-violating nature of the weak interaction that 
enables us to actually see it in experiments where electromagnetic or strong
interactions dominate.  
In the experiments considered at this meeting, 
the small contribution of the weak interaction is
extracted by ``beating'' it against the dominant electromagnetic interaction, i.e. by
taking advantage of the quantum mechanical interference.  In one type of experiment, 
the neutral weak current so extracted is compared with the
corresponding electromagnetic current to learn about hadron structure.  In the other, the
structure of the target is assumed to be known and the couplings and structure of the interaction itself are
investigated.  Let us look first at the hadronic structure physics.

The strange quark contribution to hadronic structure has been the focus of 
electron-scattering parity-violation experiments for some time.  The neutral weak current
of the nucleon is critical in this connection, because the flavor singlet current (the difference
between the electromagnetic and neutral weak currents) does
not couple to the photon as the total charge of the u, d and s quarks is zero.  Cahn and
Gilman\cite{ref:cahngilman} were the first to recognize that the weak interaction could
be useful in separating contributions from the quark flavors although they chose to
neglect the contribution of the strange quarks (by assuming charge symmetry for the proton
and neutron, they predicted the neutral weak form factors and hence the elastic scattering 
asymmetry for a given weak interaction model -- the subject of investigation at the time).
The development of experiments to investigate the strange quark vector currents derived
from several sources in the late 1980's.  The measurement of the polarized deep-inelastic
scattering structure function $g_1(x)$ by the EMC collaboration\cite{ref:EMC} led to the
suggestion\cite{ref:Jaffe} that the discrepancy in the measured Ellis-Jaffe sum rule\cite{ref:EllisJaffe}
was due to the contribution of strange quarks, in this case, to the axial vector matrix
element $\left < \bar s \gamma_\mu \gamma_5 s \right >$.  At about the same time, Steve Polluck was
re-examining the elastic e-p scattering asymmetry in his thesis\cite{ref:PolluckThesis} (he has also been a
strong advocate of the parity-violation experiment to measure the neutron radius in lead).
Perhaps the key development was the connection drawn by Kaplan and 
Manohar\cite{ref:KaplanManohar} between the axial-vector strange quark
matrix elements in deep-inelastic scattering and elastic neutrino 
scattering\cite{ref:E734} and the
suggestion that the analogous vector current matrix element should be investigated.  Thereafter the idea
was picked up by McKeown,\cite{ref:McKeown} Napolitano\cite{ref:Napolitano} and myself.\cite{ref:Beck}

Understanding the role of strange quarks in the structure of ``light quark'' hadrons is
likely to be difficult -- as was pointed out by a number of speakers at the conference.
They are neither light nor heavy; in fact their mass of $\sim 100$ MeV is roughly comparable to the QCD
scale ($\Lambda_{QCD} \sim 200$ MeV).  There are a number of indications, however, that
their role could be an important one in the structure of light quark hadrons as well as
in such diverse areas as color superconducting quark matter\cite{ref:strangeQuarkMatter} and the nature of the 
chiral phase transition in hot nuclear matter\cite{ref:phaseTransition}.

Despite the difficulties, several different approaches to modeling light hadron structure
including strange quarks were presented\footnote{See also the recent review by Beck and
Holstein\cite{ref:BeckHolstein}}.  In the model of Riska and 
collaborators\cite{ref:Riska}, constituent quarks interact via the exchange of Goldstone
bosons.  Whereas this model avoids the large couplings associated with interactions of 
Goldstone bosons with the nucleon, the meson-quark couplings must be inferred.
Further, at present, only the lowest order interactions have been included.  

It has been
noted that a widely quoted result from chiral perturbation theory, namely the 
model-independent slope of $G_M^s$ at $Q^2=0$\cite{ref:Meissner} is, in fact, model dependent
because some 1-loop order diagrams are canceled at higher order.\cite{ref:Holstein}  This problem was 
signaled by the realization that at the calculated order, the slope should simply be that
of the isoscalar magnetic moment (which it is not).  As a guide, using a
form factor with a cutoff, suggests that the actual slope might be only 1/4 - 1/3 as large
as originally predicted.  

Dispersion calculations provide perhaps the surest way to
use available experimental information together with well-defined theoretical techniques
to calculate strange quark form factors.\cite{ref:Hammer}  In principle, they combine the simpler ``pole''
and ``loop'' calculations performed previously.  One of the interesting aspects of these 
calculations is that they give the same sign for $G_E^s$ and $G_M^s$ ($<0$) resulting in
a potential conflict with the measured HAPPEX point (showing that $G_E^s + 0.39 G_M^s$ is
near zero).  It is therefore possible that such calculations should include higher lying
meson resonances (though the requisite data are sparse at best).  

Over the past couple
of years, it has been realized that generalizing the standard forward Compton scattering 
diagram used to describe electron scattering to the off-forward case leads to a set of
structure functions (Generalized Parton Distributions - GPD's) that link deep-inelastic
and elastic physics.\cite{ref:Vanderhaeghen}  For example, with the assumption of the dominance of the simple
hand-bag diagram, the Dirac form factor $F_1^s$ can be calculated as a
weighted integral of the difference between $s(x)$ and $\bar s (x)$ measured in 
deep-inelastic neutrino scattering.  The weighting itself depends on determining the
parameters of Regge trajectories.  This interesting approach suggests that if $F_1^s$
turns out to be significant ($s(x)$ and $\bar s (x)$ are nearly indistinguishable), higher
lying meson resonances may be responsible.

Is it possible to forget the models and just wait for the lattice calculations to provide
the answers?  The answer for me is clearly no.  The reason for starting this experimental
program in the first place was to help develop the picture of how the quarks (in this case
the sea quarks) and gluons are organized in the nucleon.  Simply calculating the form
factors on the lattice and comparing with data is, in my view, not physics.  
Physics requires a reasonably compact and efficient description of natural phenomena.  In
this case I expect we will be able to identify the appropriate quasi-particles to describe hadron
structure and be able
to describe the interactions of these quasi-particles in some effective theory.

In any case, despite strong progress in lattice calculations, there is still quite a long
way to go.\cite{ref:Pene}  Two prominent steps have recently been taken.  Dynamical quarks can now be
included in calculations albeit with a significant cost in computation time ($\sim$ x10).  It
does not appear to date as if there are any quantities where this contribution is dominant,
but the calculations are at an early stage.  The problem of getting down to physical quark 
masses (a few MeV for the light quarks) has been largely circumvented by taking advantage
of known mass dependences in chiral perturbation theory to extrapolate the lattice 
results to the appropriate mass.  Teraflop computers are now on the way; a few 10's of
teraflops are probably necessary to calculate strange quark form factors.  However, getting all the way to including
$s$ quarks in an exact calculation is still problematic in principle.  The methods used
to include dynamical quarks rely on having an even number of flavors.  It is not clear how
the large mass difference between the strange and charmed quarks can be handled, although
there is time to work on this problem while the computers and algorithms catch up on
dynamical light quarks.

A couple of talks at this workshop addressed the weak interactions of hadrons as
measured in parity-violation experiments.\cite{ref:Musolf}  Measurement of
the weak pi-nucleon coupling constant, called variously $h_{\pi NN}$ and $f_\pi$ in the
literature, by looking at parity-violating pion photoproduction near threshold was 
proposed.\cite{ref:Martin}  Much of 
our understanding of these phenomena is coded in the Desplanques-Donoghue-Holstein (DDH)
parameterization of the leading order coupling constants.  At present, the determination
of $h_{\pi NN}$ and the other parameters from measurements in nuclei, from $pp$ scattering
and from atomic parity-violation (via the anapole moment) are in some disarray.  I know
of at least one alternative to measuring this particular coupling -- via the $np \rightarrow d\gamma$
capture in the development stages at Los Alamos.  The experiment discussed at this meeting
is an important but difficult one which will challenge a next generation of 
electron or real photon parity-violation experiments.

The techniques that make use of the weak interaction to look at hadronic structure physics
are clearly also applicable, in the proper context, to studying the weak interaction 
itself.  Three such situations were discussed at this meeting -- all with daunting
requirements for precision.  Having said this, what is remarkable is that one of the
measurements, the weak mixing in the cesium atom, has been completed at a level which 
provides powerful constraints on TeV scale physics (making measurements of weak electric
dipole matrix elements at the tiny scale $10^{-18}$ e-cm).\cite{ref:Bouchiat}  In fact, the 0.3\% experimental
accuracy for the weak charge of the cesium nucleus 
has outstripped the atomic calculations necessary to relate the observable
to the underlying theory.  Work continues to improve these calculations.  At this meeting
we heard a very promising report on the M\"oller scattering parity-violation experiment
(SLAC E158).\cite{ref:Souder}  In this case the physics asymmetry is a bit more than $10^{-7}$ and the goal for
the overall uncertainty is about $10^{-8}$.  Remarkably, with a lot of hard work on
controlling systematic uncertainties associated with the beam, as well as on backgrounds,
this experiment is poised for its main data-taking phase.  Lastly, the approved Jefferson
Lab experiment, Qweak, will measure the proton weak charge in parity-violating
electron-proton scattering at very low momentum transfer ($Q^2=0.03$ GeV$^2$).\cite{ref:Pitt}  Again,
with a desired statistical precision of $10^{-8}$, this will be a difficult measurement.
The aim is to run this experiment in 2005.

These experiments are particularly sensitive to certain types of new physics.\cite{ref:Pitt}  Heavier 
$Z^{\hspace{0.1em}\prime}$ gauge bosons are more visible at low energies than they are at the ordinary $Z$ pole
where the most precise electroweak studies have been done.  There are two schemes for
mixing families, R-parity violating supersymmetry and its cousin -- leptoquarks -- wherein
quarks are transformed into leptons and vice versa, which could make significant
contributions to these observables.  One of the nice features of this set of measurements
is that these new physics effects have complementary impacts.  For example, R-parity
violating SUSY has a larger impact in Qweak than E158 and vice versa for leptoquarks.
This discussion brings to mind a recent speech by the US Presidential Science Advisor and
former Brookhaven director John Marburger in which he challenged us to imagine doing
physics at the frontier after the last large accelerator is built.  His speculation is
that a significant part of our work will involve precision measurements of which these
experiments may be precursors.

\section {Open Questions}

There are a number of ``technology'' issues that are apparent in our current state of the 
art.  Although progress has been made, successful completion of the first analysis (probably E158)
where the systematic uncertainties are in the few ppb range will be a milestone.  This
will likely require continued work on control of helicity-correlated beam properties
associated particularly with strained crystals as discussed at this meeting.  Similarly,
successful completion of an analysis associated with more sophisticated detectors
(e.g. PVA4), where some of the systematic considerations are different, will also be a
milestone.  There have frequently been questions about how accurately the strange quark
contributions to the form factors should be determined -- I would say that the level of a
few \% of the overall form factor, as expected in the G0 experiment, is sufficient to 
significantly advance the picture of their role in hadron structure.  At present there
do not seem to be particular values commonly emerging from theory, except perhaps for the
clustering of a number of results for $\mu_s$ around -0.3 -- precision of a few \% is clearly
enough to evaluate this prediction.  Finally, there is the old
problem of charge symmetry in the proton-neutron system.  We really need to know whether
the probability of finding a $u$ quark at radius $r$ in the proton is the same as finding
a $d$ quark at the same radius in the neutron.  Gerry Miller has done a calculation\cite{ref:Miller}
that indicates the effects in the strange quark experiments will be very small -- directly
applicable experimental input is scarce or non-existent.

In terms of physics, I would like to address four topics.  First,
there has been progress in modeling strange quark effects, notably in the dispersion 
theory approaches and in continuing exploration of new connections such as generalized
parton distributions.  However, as the experiments push forward, it is important to try to picture
what happens to these effects as the momentum transfer increases toward 1 GeV$^2$.  
Although a number of technical problems have been solved in lattice calculations, it
seems we are still quite some distance from reliable strange quark matrix elements.

Second, I think the axial current observed in the SAMPLE experiment is interesting to
follow up.  Although we have had several calculations from Mike Ramsey-Musolf and collaborators,
I believe there is room for others to contribute here.  First of all, perhaps it would
be useful to focus on the observable axial current difference between neutrino and electron
scattering as the definition of what to calculate (or approximate) in order to relieve
confusions about definitions.  It might be beneficial to follow the lead of Claude Bouchiat
whose picture of nucleons with longitudinal (relative to momentum) components of spin,
arising from weak interactions between nucleons, underlies the anapole moment effects
calculated for nuclei and observed in the atomic cesium experiment.\cite{ref:Bouchiat,ref:CBouchiat}
  I expect we will find
in the end that questions of definition will recede and that this observable will tell us
about weak interactions among the quarks in the nucleon.

Third, and on a closely related note, continuing to press on calculations of (and perhaps
some related measurements relevant to) all the radiative, or perhaps
one should say higher-order corrections, will be critical to continuing to make progress
across this broad field.  Related effects enter in a number of other important 
measurements -- neutron $\beta$ decay and muon $g-2$, for example, and we are learning (again)
about their critical role to interpretation of the experiments.  Of course, even in fundamental measurements,
it is again effects
associated with hadrons or quark loops that are the toughest to address.

Lastly, perhaps the most common question we hear about the strange quark effects is ``what
if you see zero?''  On the surface this may seem like a defeat, but I believe such a result
is, if anything, more directly informative than measuring some particular values.  I think
it is important to discuss the physics of this possibility more vigorously and I
particularly like the framework suggested by von Harrach in our discussions at this meeting.
To start, even though $s$ quarks may contribute to the nucleon mass and spin (where quarks
and antiquarks contribute with the same sign), they may be relatively invisible in the
(charge conjugation even) observables like the charge and magnetization distributions.  The
issue, then, is whether the $s$ quarks and anti-quarks that are there either a) are
significantly separated or b) are in a relative spin singlet (such that their magnetic
moments are parallel).  The first possible cause of a zero or near zero result is that the $s$
and $\bar s$ simply do not live long enough to interact with the medium and become significantly
separated.  This may be in contrast to the explanation for the excess of $\bar d$ quarks
over $\bar u$ quarks in the proton (Gottfried sum and direct Drell-Yan data). In this case, there
are as many fervent defenders of the pion cloud explanation (where the pairs do interact
and separate into some quasi-nucleon and pion packages) as there are of some effect related
to the Pauli principle (where simply fewer $u \bar u$ pairs are formed).  The recent
observations of the HERMES collaboration that the spin carried by anti-quarks is very
small at high energies suggests that in this case the quarks are produced in relative
spin-singlet states (gluon spin carried by relative orbital angular momentum between the
$q$ and the $\bar q$), likely a simple result of helicity conservation.  Therefore, in the second
case, it may be that the soft physics is sufficiently different that the pairs are created in spin
triplet states leading to a zero contribution to the magnetization.  These questions,
I think, remain among the most interesting in nucleon structure physics.

\section*{Acknowledgments}
On behalf of all the conferees I would like to thank the Johannes Gutenberg 
Universit\"a t Mainz, the Institut f\"u r Kernphysik and its director Thomas Walcher for 
their kind hospitality in hosting the conference, the staff of the Institut for the
organization and support that allowed the conference to run so smoothly, the students
of the Institut who did all the real work (and indeed the grad students behind the scenes of
all the presentations).
Frank Maas, the head of the local organizers, deserves a special vote of thanks for
organizing a very stimulating and enjoyable week.


\begin{thebibliography}{0}

\bibitem{ref:Aulenbacher} K. Aulenbacher (Mainz), talk at this workshop.

\bibitem{ref:Humensky} B. Humensky (Virginia), talk at this workshop.

\bibitem{ref:Baylac} M. Baylac (JLab), talk at this workshop.

\bibitem{ref:Farkhondeh} See, for example, M. Farkhondeh (MIT-Bates), talk at this workshop.

\bibitem{ref:Grames} J. Grames (JLab), talk at this workshop.

\bibitem{ref:Burtin} E. Burtin (Saclay), talk at this workshop.

\bibitem{ref:Collin} B. Collin (Orsay), talk at this workshop.

\bibitem{ref:Pitt} M. Pitt (Virginia Tech), talk at this workshop.

\bibitem{ref:Michaels} R. Michaels (JLab), talk at this workshop.

\bibitem{ref:Furget} C. Furget (Grenoble), talk at this workshop.

\bibitem{ref:Souder} P. Souder (Syracuse), talk at this workshop.

\bibitem{ref:Maas} F. Maas (Mainz), talk at this workshop.

\bibitem{ref:cahngilman} R. Cahn and F. Gilman, {\it Phys. Rev.}
{\bf D17}, 1313 (1978).

\bibitem{ref:EMC} J. Ashman, et al. (European Muon Collaboration), {\it Nucl. Phys.} {\bf B328} 1 (1989). 

\bibitem{ref:Jaffe} R. Jaffe, {\it Phys. Lett.} {\bf B193} 101 (1987).

\bibitem{ref:EllisJaffe} J. Ellis and R. Jaffe, {\it Phys. Rev.} {\bf D9} 1444 (1974).

\bibitem{ref:PolluckThesis}S. Polluck, {\it Nucl. Phys.} {\bf A461} 553 (1987).

\bibitem{ref:KaplanManohar}D. Kaplan and A. Manohar, {\it Phys. Lett.} {\bf B310}, 527 (1988).

\bibitem{ref:McKeown} R. McKeown, {\it Phys. Lett.} {\bf BB219}, 140 (1989).

\bibitem{ref:Napolitano} J. Napolitano, {\it Phys. Rev.} {\bf C43}1473 (1991).

\bibitem{ref:Beck} D. H. Beck, {\it Phys. Rev.} {\bf D39}, 3248 (1989).

\bibitem{ref:E734} L.A. Ahrens, {\it Phys. Rev. Lett.} {\bf 35}, 785 (1987).

\bibitem{ref:strangeQuarkMatter} M. Alford, J. Berges and K. Rajagopal {\it Nucl.Phys.} {\bf B558} 219 (1999). 

\bibitem{ref:phaseTransition} See, for example, J. Pel\' aez, hep-ph/0205124.

\bibitem{ref:BeckHolstein} D. Beck and B. Holstein, {\it Int. J. Mod. Phys.} {\bf E10}, 1 
(2001).

\bibitem{ref:Riska} D. Riska (Helsinki), talk at this workshop.

\bibitem{ref:Meissner} T.R. Hemmert, B. Kubis, and U.-G. Meissner,
{\it Phys. Rev.} {\bf C1260}, 045501 (1999).

\bibitem{ref:Holstein} B. Holstein (Massachusetts), talk at this workshop.

\bibitem{ref:Hammer} H. Hammer (Ohio State), talk at this workshop.

\bibitem{ref:Vanderhaeghen} M. Vanderhaeghen (Mainz), talk at this workshop.

\bibitem{ref:Pene} O. Pene (Paris), talk at this workshop.

\bibitem{ref:Musolf} M. Ramsey-Musolf (Caltech), talk {\it Parity-violating low energy constants} at this workshop.

\bibitem{ref:Martin} J. Martin (Caltech), talk {\it Parity-violating low energy constants} at this workshop

\bibitem{ref:Bouchiat} M. Bouchiat (ENS), talk at this workshop.

\bibitem{ref:Miller} G. Miller, {\it Phys. Rev.} {\bf C57}, 1492 (1998).

\bibitem{ref:CBouchiat} C. Bouchiat and C. Piketty, {\it Z. Phys.} {\bf C49}, 91 (1991).

\end{thebibliography}
\end{document}